\documentclass[12pt]{iopart} %

\usepackage{iopams}
\usepackage{graphicx,subfigure}
\usepackage{epstopdf}
\usepackage{pdfcomment}

\begin{document}

\title{Lattice dynamics and thermophysical properties of h.c.p. Re and Tc from the quasi-harmonic approximation}

\author{Mauro Palumbo$^1$, Andrea Dal Corso$^{1,2}$}
\address{$^1$International School for Advanced Studies (SISSA), Trieste, ITALY}
\address{$^2$CNR-IOM, Trieste, ITALY}
\ead{mpalumbo@sissa.it}

\begin{abstract}
  We report first-principles phonon frequencies and anharmonic thermodynamic properties of h.c.p. Re and Tc calculated 
  within the quasi-harmonic approximation, including Gr\"uneisen parameters, temperature-dependent lattice parameters, thermal expansion, 
  and isobaric heat capacity.
  We discuss 
  the differences between a full treatment of anisotropy and a simplified approach with a constant $c/a$
  ratio. 
  The results are systematically compared with the available experimental data and an overall satisfactory agreement is obtained.
\end{abstract}

%
%

\maketitle   

\section{Introduction}
Rhenium (Re) and technetium (Tc) belong to the same group in the periodic table and both present experimental difficulties
in the determination of their thermophysical properties, particularly at high temperature. The former
is a transition metal relatively rare in the Earth crust with one of the highest melting points (3458~K). Among refractory elements, Re is
unique in presenting an hexagonal-close-packed (h.c.p.) crystal structure contrary to other body-centered-cubic elements such as W, Mo and Ta.
Because of its different crystal structure, Re is the only refractory element known to have relatively high ductility at low temperature. 
Despite its high cost, it has found applications as a high pressure gasket material~\cite{87Sch} and in Ni-base superalloys~\cite{01Rae}.
Although it is a minor component of these alloys, which contain up to 15 elements, its influence on their final properties is remarkable. 
Nonetheless, there have been limited experimental investigations on h.c.p. Re. Measurements of phonon frequencies are difficult because of the high 
neutron absorption rate, while measuring its thermophysical properties requires equipments suitable for high temperatures. 
Tc, first discovered by Perrier and Segre in 1937, is also an h.c.p. transition metal but it does not have a stable isotope.
Nonetheless, the short-lived $\gamma$-emitting isotope $^{99m}$Tc is used in diagnostic nuclear medicine, while the $\beta$-emitting $^{99}$Tc is 
obtained in nuclear reactors and represents an important challenge in nuclear waste management. As for Re, experimental data
are scarce because of intrinsic difficulties in measurements.

These experimental difficulties make it even more important to have theoretical predictions of the properties of these elements since
the thermophysical properties of pure elements are important in a multi-scale 
modeling approach which aims at describing the properties of complex materials and systems
from the description of the pure elements~\cite{14Pal}.

Two experimental studies of Re phonon frequencies have been reported in the literature~\cite{82Wak,86Shi}. 
Wakabayashi et al. report phonon frequencies along the $\Gamma$ $\to$ A direction, while Shitikov et al. report the
phonon density of states (DOS). Experimental data of thermophysical properties are presented 
in~\cite{96Arb,86Pot,66Con,60Leh,61Tay,62Rud,53Smi,65Con,53Hor} and 
references therein. They include thermal expansion and heat capacity, but data are rather limited compared to other transition metals. 
Theoretical studies 
of pure Re are also available but mostly limited to 0 K~\cite{88Wat,95Fas,99Ste,12Jon,paperAndrea}. 
Phonon dispersions at different pressures (0, 50 and 100 GPa) were calculated in~\cite{12Lv} in the harmonic approximation. These authors, 
however, did not compare their results with the experimental phonon frequencies in~\cite{82Wak,86Shi} and did not calculate anharmonic quantities.
The available experimental and theoretical data for Tc were reviewed by Rard et al.~\cite{99Rar}. Wakabayashi et al.~\cite{82Wak} 
reported experimental phonon frequencies at room temperature along several directions in the Brillouin zone (BZ).
A few conflicting heat capacity  data sets were reported by Shirasu and Minato~\cite{02Shi}, Boucharat~\cite{97Bou} and Spitsyn et al.~\cite{75Spi}
above room temperature. Some theoretical estimates of the heat capacity and other quantities from different investigators are summarized by 
Rard et al.~\cite{99Rar}. A recent paper by Weck and Kim~\cite{15Wec} reports theoretical quasi-harmonic calculations using the Birch-Murnaghan and Vinet 
equations of state. Souvatzis et al.~\cite{08Sou} have also done phonon and quasi-harmonic calculations on h.c.p. Tc using a supercell method and
including the evaluation of the anisotropic thermal expasion.

In the present work we present anharmonic results for h.c.p. Re and Tc calculated in the quasi-harmonic approximation using the Quantum Espresso package~\cite{QEweb,09Gia} and the 
recently developed thermo\_pw code~\cite{thermopw}. Phonon dispersions were computed
as a function of volume and lattice parameters. From these, Gr\"{u}neisein parameters, temperature-dependent thermal expansions 
and other thermophysical quantities were computed. The results were first calculated
assuming a constant $c/a$ ratio and then on a grid of ($a$,$c/a$) values that allow a full treatment of the anisotropic thermal expansion tensor. 
A systematic comparison with experimental data and previous results is presented.

\section{Computational method}\label{sec:Compmethod}

All calculations in the present work were carried out using Density Functional Theory (DFT) and plane waves basis sets as implemented in the 
Quantum Espresso (QE) package~\cite{QEweb,09Gia}. The thermo\_pw package~\cite{thermopw} was used to run simultaneously and asynchronously 
many self-consistent and phonon calculations with an additional level of parallelization with respect to QE. 
The Local Density Approximation (LDA)~\cite{81Per}, the Perdew-Burke-Ernzerhof Generalized Gradient Approximation (PBE-GGA)~\cite{96Per} 
and the revised PBE functional for densely-packed solids (PBEsol)~\cite{08Per} were tested for the exchange-correlation functional. We employed the
Re.pz-spn-kjpaw\_psl.1.0.0.UPF, Re.pbe-spn-kjpaw\_psl.1.0.0.UPF and Tc.pz-spn-kjpaw\_psl.1.0.0.UPF pseudopotentials from the pslibrary1.0~\cite{pslibrary,14Dal}.
The re\_lda\_v1.2.uspp.F.UPF from the Garrity, Bennett, Rabe and Vanderbilt (GBRV) library~\cite{14Gar} and Re.pbesol-spn-kjpaw\_psl.1.0.0.UPF from 
the pslibrary1.0 were also used for a few comparisons.

Computational settings were chosen to ensure that all investigated properties and in particular phonon frequencies are well converged. 
The kinetic energy cut-off was set to 50 Ry while the charge density cut-off was set to 300 Ry for Re PBE/PBEsol and to 200 Ry for all others. 
The integration over the Brillouin Zone (BZ)
was  performed employing 15$\times$15$\times$10 (for Re) and 18$\times$18$\times$12 (for Tc) {\bf k}-points Monkhorst-Pack meshes 
and a Methfessel-Paxton smearing scheme has been used with 0.02 Ry width.
Phonon frequencies were computed using Density Functional Perturbation Theory (DFPT), as implemented in the phonon code~\cite{01Bar}, on a 
6$\times$6$\times$4 grid of {\bf q}-points and Fourier interpolated in the BZ.

In the quasi-harmonic approximation~\cite{10Bar}, the Helmholtz energy of a crystalline solid is assumed to be 
\begin{equation}
F(T,X) = U_0(X)+F^{\rm{vib}}(T,X)+F^{\rm{el}}(T,X),
\label{eq:F}
\end{equation}
where $U_0$ is the static energy at 0~K, $F^{\rm{vib}}$ the contribution of lattice vibrations and $F^{\rm{el}}$ the energy due to electronic
excitations.  In the adiabatic approximation, each term is treated separately.
X refers to any parameter upon which the above energies may depend. In the most common case, i.e. for
 cubic lattices, X is the unit cell volume. In general, it may refer to internal distortions and/or the strain tensor. In the case of an h.c.p.
cell, X refers to the two lattice parameters $a$ and $c/a$, although it is very common to assume that $c/a$ is constant
as a function of temperature and thus consider the variation of $F$ only as a function of $a$ (or equivalently $V$). 
For a given $X$, the vibrational Helmholtz energy per cell is calculated as in the harmonic approximation: 
\begin{equation}
 F^{\rm{vib}}(X,T) =\frac{1}{2 N} \sum_{\vec{q},\nu}{\hbar \omega(\vec{q},\nu,X)} + 
 \frac{k_{\rm B} T}{N} \sum_{\vec{q},\nu}\ln[1-\exp(\frac{-\hbar \omega(\vec{q},\nu,X)}{k_{\rm B} T})].
\label{eq:Fharm}
\end{equation}
The first term on the right-hand side of the above equation is the Zero-Point Energy (ZPE), the second term is the phonon contribution at finite
temperatures. The sums in Eq.~(\ref{eq:Fharm}) are taken over the phonon frequencies $\omega(\vec{q},\nu,X)$, where $\nu$ denotes the different
phonon branches and $\vec{q}$ the wave vectors within the first BZ. $k_B$ is the Boltzmann constant, $\hbar$ the reduced Planck constant, 
$T$ the absolute temperature, and $N$ the number of unit cells in the solid.

In the following we carried out calculations both (a) assuming a constant $c/a$ and (b) evaluating the Helmholtz energy on a full grid ($a$, $c/a$).
In the former case, we computed the phonon frequencies at 9 different volumes (or equivalently $a$ values), keeping $c/a$ at the 0~K 
equilibrium value and fitting the resulting energies using a Murnaghan equation of state (EOS)~\cite{44Mur}. In the latter case, the calculations
were carried out on a grid of points of $a$ and $c/a$ and then fitted with a quartic polynomial. The treatment of anisotropic quantities 
for the h.c.p. lattice is detailed in the following section, while in the rest of this section we assume $X=V$.
The minimum $U_0(V)+F^{\rm{vib}}(T,V)$ at each temperature corresponds to zero pressure. In this way, the temperature dependence
of the free energy $F$, the volume $V$ and the isothermal bulk modulus $B_T$ is directly obtained from the Murnaghan EOS. 
By numerical derivation 
of the volume we obtained the volume thermal expansion coefficient $\beta$ according to its definition
\begin{equation}
 \beta =\frac{1}{V} \left(\frac{\partial V}{\partial T}\right)_{P}.
\label{eq:beta}
\end{equation}
The so-called Gr\"{u}neisein mode parameters were calculated as
\begin{equation}
\gamma_{{\bf q},\nu} = - {V\over \omega(\vec{q},\nu,V)} {\partial
\omega(\vec{q},\nu,V) \over \partial V} ,
\label{eq:Grun}
\end{equation}
from the computed frequencies. 
The constant volume (isochoric) heat capacity was obtained from the phonon frequencies calculated as
in the harmonic approximation
\begin{equation}
 C_{V} =\frac{k_{\rm B}}{N} \sum_{\vec{q},\nu}\left(\frac{\hbar\omega(\vec{q},\nu)}{k_BT}\right)^2 \frac{\exp(\hbar\omega(\vec{q},\nu)/k_B T)}{\left[\exp(\hbar\omega(\vec{q},\nu)/k_B T)-1 \right]^2}.
\label{eq:Cv}
\end{equation}
at each fixed volume in the quasi-harmonic grid. The isochoric heat capacity at each temperature is then obtained 
interpolating at the temperature-dependent volume values obtained at each temperature from the minimization of the free energy.

Finally, the constant pressure (isobaric) heat capacity was obtained as~\cite{72Wal}
\begin{equation}
 C_{P} = C_{V} + TV\beta^{2}B_{T} .
\label{eq:Cp}
\end{equation}
The electronic contribution to the heat capacity was obtained in the single-particle approximation from the electronic DOS 
at the equilibrium lattice parameters at 0~K, as described in details in Ref.~\cite{14Pal,99Gri}.

\section{Anisotropy in hexagonal structures}\label{sec:anisotropy}
In non-cubic systems a full treatment of anisotropy should consider the variable $X$ in Eq.~\ref{eq:F} as 6 independent components of the  
symmetric $3\times 3$ strain tensor $\epsilon$ given by
\begin{equation}
{\bf a}'_{j,i} = {\bf a}_{j,i} +  \sum_{k=1}^3 \epsilon_{j,k} {\bf a}_{k,i},
\end{equation}
where the three primitive vectors ${\bf a}_1$, ${\bf a}_2$, and ${\bf a}_3$ refer to the unperturbed solid while
${\bf a}'_1$, ${\bf a}'_2$, and ${\bf a}'_3$ refer to the strained solid.
Similarly, we need to consider the $3\times 3$ stress tensor $\sigma$. 
From the elasticity theory, we have that, for small strains, stress is proportional to strain 
\begin{equation}
\sigma_{i,j} =\sum_{lm} C_{i,j,l,m} \epsilon_{l,m} ,
\end{equation}
where $C_{i,j,l,m}$ is the 9$\times$9 elastic constants tensor.
Since the stress and strain tensors are symmetric, it is common to introduce the Voigt notation which reduces the number of components of the
stress and strain tensors to 6 and those of the elastic constants to 21 (in the triclinic system). For example the strain tensor
in the Voigt notation is given by $\epsilon_1=\epsilon_{1,1}$, $\epsilon_2=\epsilon_{2,2}$, $\epsilon_3=\epsilon_{3,3}$, 
$\epsilon_4=2 \epsilon_{2,3}$, $\epsilon_5=2 \epsilon_{1,3}$, and $\epsilon_6=2 \epsilon_{1,2}$.

The definitions of thermodynamic functions such as internal energy, enthalpy, Helmholtz and Gibbs energies can be generalized for a strained solid ~\cite{72Wal}. 
For example, the anisotropic Helmholtz energy is given by
\begin{equation}
dF=-SdT + V \sum_{i=1}^6 \sigma_i d\epsilon_i.
\end{equation}
At a given temperature T we have an (anisotropic) equation of state as 
\begin{equation}
\sigma_i= {1\over V } \left({\partial F \over \partial \epsilon_i}\right)_T ,
\label{Feos}
\end{equation}
and the equilibrium strains at $\sigma=0$ are obtained minimizing the free energy.
In the anisotropic case, Murnaghan and other volume-related EOS 
are not applied and the above equation can be expressed analytically using quadratic and quartic polynomials.
The thermal expansion is a tensor given by
\begin{equation}
\alpha_{i} = \left({\partial \epsilon_{i}(T) \over  \partial T }\right)_\sigma,
\label{sigmaaniso}
\end{equation}
where the $\epsilon_{i}$ are the already introduced strain components in the Voigt notation.
Temperature-induced strains do not change the shape of the unit cell, hence only certain
kinds of strain are possible, depending on the symmetry of the system.
The components of the strain tensor in Eq.~\ref{sigmaaniso} can also be obtained from the temperature variation of the lattice parameters. In the present hexagonal case, the above general equations 
are simplified by symmetry and only two thermal expansion terms
are independent, i.e. $\alpha_{\perp}=\alpha_{1}=\alpha_{2}$ and $\alpha_{\parallel}=\alpha_{3}$.
In practice we derive the variation with temperature of $a$ and $c$ (or equivalently $c/a$) from the minimization of $F$.
From these lattice parameters as a function of temperature, the hexagonal thermal expansions
were obtained as:

\begin{equation}
\alpha_1= \alpha_2 = {1\over a(T) } { d a(T) \over d T },
\label{alphaa}
\end{equation}

\begin{equation}
\alpha_3= {1\over c(T) } {d c(T) \over d T}.
\label{alphac}
\end{equation}

Finally, the volume thermal expansion $\beta$ is given by $\beta=2 \alpha_1 + \alpha_3$.

For Re, a total of 45 grid points ($9\times5$) were used with steps of 0.01 a.u. and 0.005 in $a$ and $c/a$, respectively. The grid was centered 
at the equilibrium values of $a$ and $c/a$ at 0~K.
The resulting total energies at each grid point
were fitted with a quartic polynomial as a function of $a$ and $c/a$. 
Similarly for Tc, a $5\times5$ grid centered at the equilibrium values of $a$ and $c/a$ at 0~K was chosen, 
with steps $\Delta a$=0.01 a.u. and $\Delta (c/a)$=0.003. These grid steps were set up with special care in order to avoid imaginary frequencies 
for the highest $(a,c/a)$ values and at the same time to span the largest possible parameters space.

\section{Results}

\subsection{Rhenium}\label{secRe}

We first minimize the total energy at 0~K using the thermo\_pw code on a full grid of points $a$ and $c/a$ with both the LDA and PBE 
exchange-correlation functionals. The equilibrium crystal parameters are reported in Table~\ref{tab:0K} together with previous results. The agreement with
previous results and experiments is satisfactory. 

\begin{table}[htp]
\caption{\label{tab:0K} Comparison between structural parameters of h.c.p. Re at 0~K from different sources. NCPP is for norm-conserving
pseudopotentials, USPP is for ultra-soft pseudopotentials, FLMTO is for full potential linear muffin-tin orbitals, FLAPW is for full potential linearized augmented plane waves. 
The PAW LDA and PBE results in this work were obtained using the Re.pz-spn-kjpaw\_psl.1.0.0.UPF and Re.pbe-spn-kjpaw\_psl.1.0.0.UPF 
pseudopotentials from the pslibrary1.0~\cite{pslibrary} and the USPP LDA results using re\_lda\_v1.2.uspp.F.UPF from the GBRV library~\cite{14Gar}.}
\begin{center}
\begin{tabular}{@{}lcccc}
 Method & $a$ & $c$ & $c/a$ & $V$  \\
        & $\AA$ & $\AA$ & & $\AA^3$ \\
PAW, LDA (0~K, this work) & 2.735 & 4.412 & 1.613 &  14.29 \\
PAW, PBE (0~K, this work) & 2.772 & 4.479 & 1.616 & 14.90 \\
PAW, LDA (298~K, this work) & 2.740 & 4.425 & 1.615 &  14.39 \\
PAW, PBE (298~K, this work) & 2.778 & 4.489 & 1.616 &  15.00\\
USPP, LDA (0~K, this work) & 2.735 & 4.412 & 1.613 &  14.29 \\
PAW, LDA~\cite{12Jon}    & 2.741 & 4.422 & 1.613 & 14.39 \\
NCPP, PBE~\cite{12Lv}   & 2.762 & 4.442 & 1.608 & 14.67 \\
NCPP, LDA~\cite{12Lv}   & 2.756 & 4.437 & 1.61  & 14.58 \\
FLMTO, LDA~\cite{95Fas}  & 2.748 & 4.474 & 1.628 & 14.62 \\
FLMTO, LDA~\cite{99Ste}  & 2.750 & 4.442 & 1.628 & 14.54 \\ 
FLMTO, PBE~\cite{99Ste}  & 2.794 & 4.513 & 1.628 & 15.25 \\
Exp. (X-ray at RT)~\cite{87Voh}    & 2.762 & 4.455 & 1.613 & 14.71 \\
Exp. (X-ray at RT)~\cite{70Liu}    & 2.761 & 4.456 & 1.614 & 14.70 \\
\end{tabular}
\end{center}
\end{table}

\begin{figure}[htb]
\includegraphics[width=\linewidth]{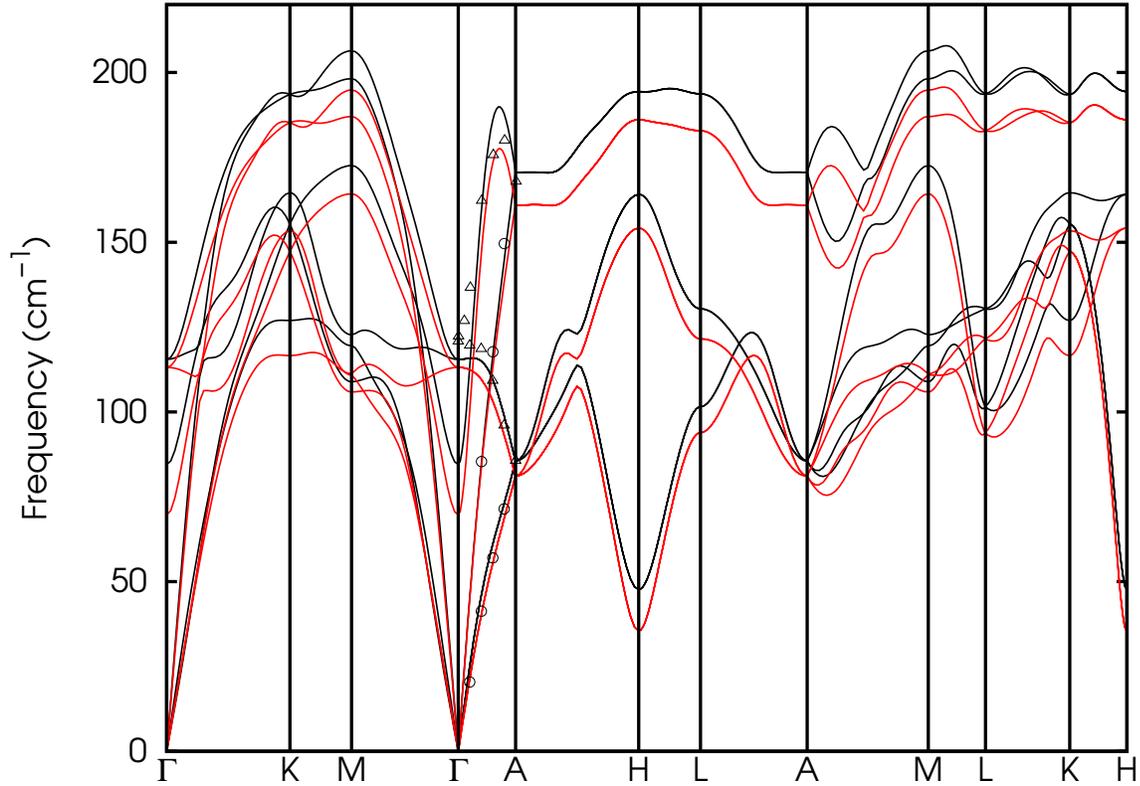}
\caption{Phonon dispersion curves along high symmetry lines in the BZ calculated with both 
LDA (black line) and PBE (red line) exchange-correlation functionals. 
The calculation was carried out at the equilibrium lattice parameters at 298~K.
Experimental data (points) are from Ref.~\cite{82Wak} at 298~K and
are limited to the $\Gamma$-A direction.
}\label{fg:phdisp}
\end{figure}

\begin{figure}[htb]
\includegraphics[width=\linewidth]{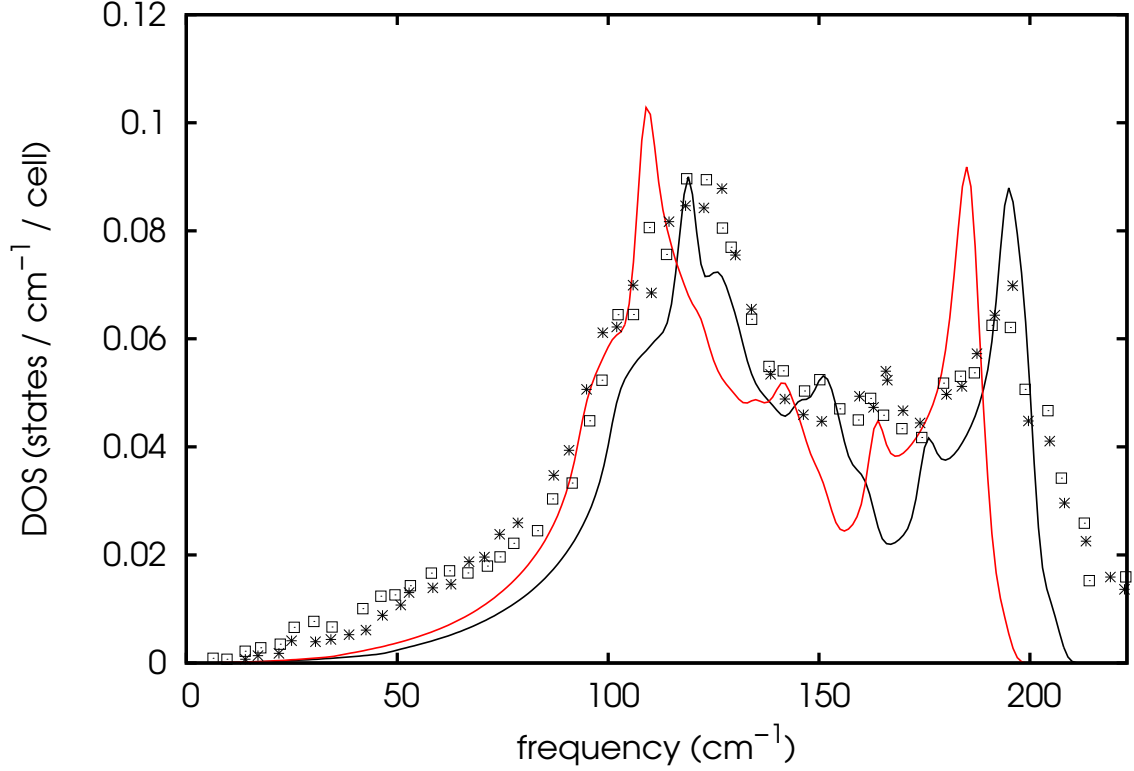}
\caption{Phonon DOS calculated with both LDA (black line) and PBE (red line) functionals. The calculation was carried out at the equilibrium lattice parameters at 298~K.
Experimental data are from Ref.~\cite{86Shi} at T=298~K (squares) and T=495~K (stars).}\label{fg:phdos}
\end{figure}

\begin{figure}[htp]
\includegraphics[width=\linewidth]{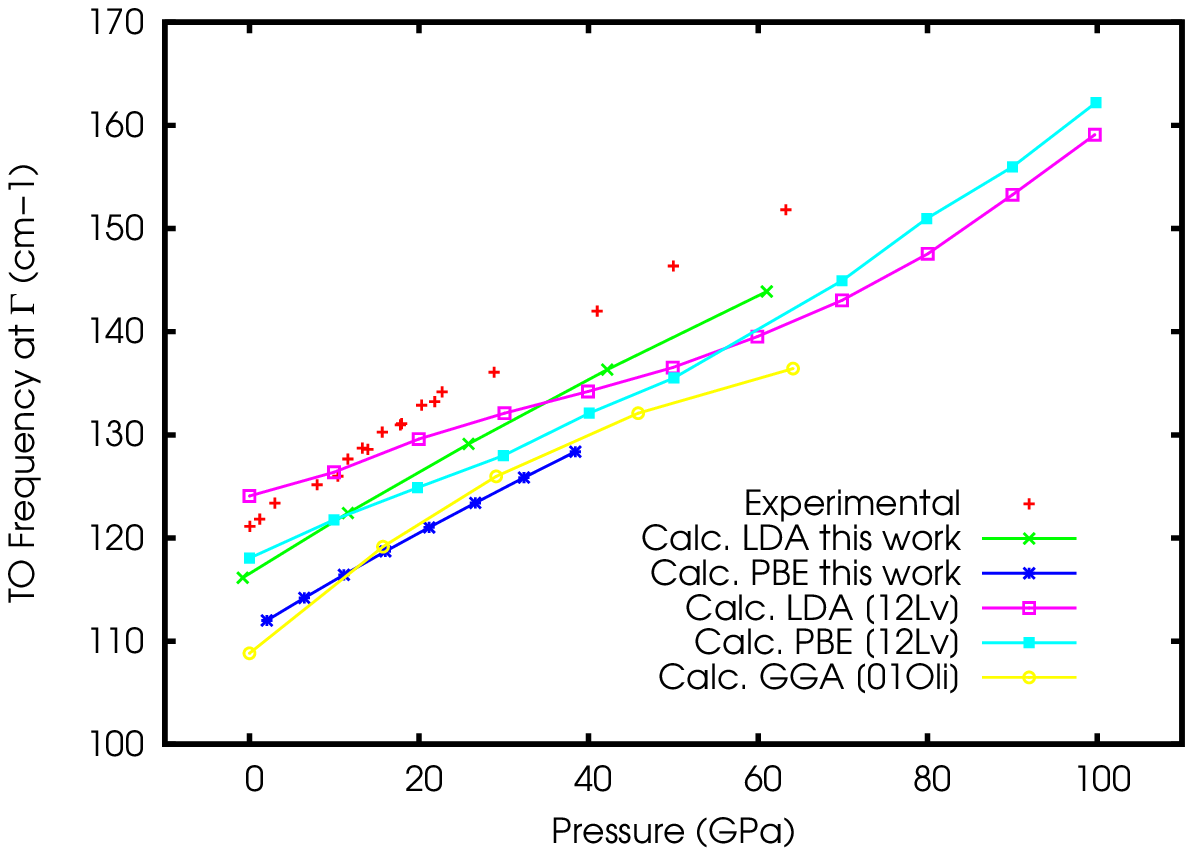}
\caption{Pressure shift of the TO phonon frequency at $\Gamma$ with both LDA and PBE calculated with the Murnaghan EOS keeping a constant $c/a$ ratio.
Experimental data were obtained from Raman spectroscopy in Ref.~\cite{01Oli}. Previous calculations from 
(12Lv)~\cite{12Lv} and (01Oli)~\cite{01Oli} are also shown for comparison.}\label{fg:freqvsP}
\end{figure}

The calculated LDA and PBE phonon dispersion curves, at the equilibrium volume at T=298~K ($a$=2.740 \AA, $c/a$=1.615 for LDA; $a$=2.778 \AA, $c/a$=1.616 for PBE), 
along high symmetry lines in the BZ are shown in Fig.~\ref{fg:phdisp} and compared with the available 
experimental data which are limited to the $\Gamma$(0,0,0) $\to$ A(0,0,1/2) direction (here and in the following the coordinates of the points in the
reciprocal lattice are in units of $2\pi/a$). The agreement with the experimental data is satisfactory, with the LDA frequencies slightly better than the PBE ones
and as expected higher.
The degeneracy of the optical frequencies at  
$\Gamma$ in the experimental data is accidental, since it is not imposed by symmetry and it is not reproduced in most other h.c.p.
elements as well as in our calculations. The significant dip of the LO frequencies from A to $\Gamma$ is also unique to Re (and Tc) and
not present in other h.c.p. metals. It is also reproduced by our calculations.
We finally remark that in the calculation this dip becomes even more pronounced when increasing the unit cell volume for example as an effect of thermal 
expansion. 
A similar softening is also observed at the point H(2/3,0,1/2). 
Interestingly, this behavior is opposite to what was found by Wakabayashi et. al.~\cite{82Wak} in experiments for Tc. 
According to these experimental results, an increase in temperature (and thus volume) in Tc results in an increase in the LO frequency 
at $\Gamma$ for Tc of more than 33 cm$^{-1}$. Unfortunately, these data are not available for Re.  
To further investigate
this effect, we carried out calculations at the $\Gamma$ point, at fixed volume and using the Fermi-Dirac distribution as smearing function 
with a smearing factor corresponding to T=300~K and T=600~K. Our results show an increase in the LO frequency at $\Gamma$ as 
89.8, 91.9 and 95.5 cm$^{-1}$ at temperatures 0, 300 and 600~K, respectively. The increasing trend is in agreement with experiments but 
largely underestimates the magnitude of experimental results for Tc. On the contrary, the effect of temperature on the frequency of the 
TO mode at $\Gamma$ is negligible.

Phonon DOS data at two temperatures (295 and 495~K) are reported in Ref.~\cite{86Shi} and compared with the calculated results in Fig.~\ref{fg:phdos}.
We note that the difference between the experimental results at the different temperatures is low and comparable with the scatter among
experimental data at a single temperature.
The difference between the calculated results (both LDA and PBE) and the experiments in the acoustic (low energy)
part of the spectrum is difficult to explain and is in contrast with the good 
agreement shown in Fig.~\ref{fg:phdisp} in the acoustic branches along the $\Gamma$-A direction. 
The PBE frequencies are in better agreement with the experiments than the LDA ones in the acoustic part of the spectrum, whereas the agreement is worse
at the highest optical frequencies. 

Experimental data of the TO phonon frequency at the $\Gamma$ point as a function of pressure obtained using Raman spectroscopy are available up to 60~GPa.
Fig.~\ref{fg:freqvsP} reports a comparison of the calculated TO phonon frequency and experimental ones.
The calculated points were obtained from a 0~K calculation at fixed volume, constant $c/a$ and the Murnaghan EOS was used to establish the $P(V)$ relationship. 
The pressure dependence of the experimental data is well reproduced by both LDA and PBE
results, but the absolute calculated values underestimate the experimental results by $\approx$4\% and $\approx$8\%, respectively.

\begin{figure}[htb]
\includegraphics[width=\linewidth]{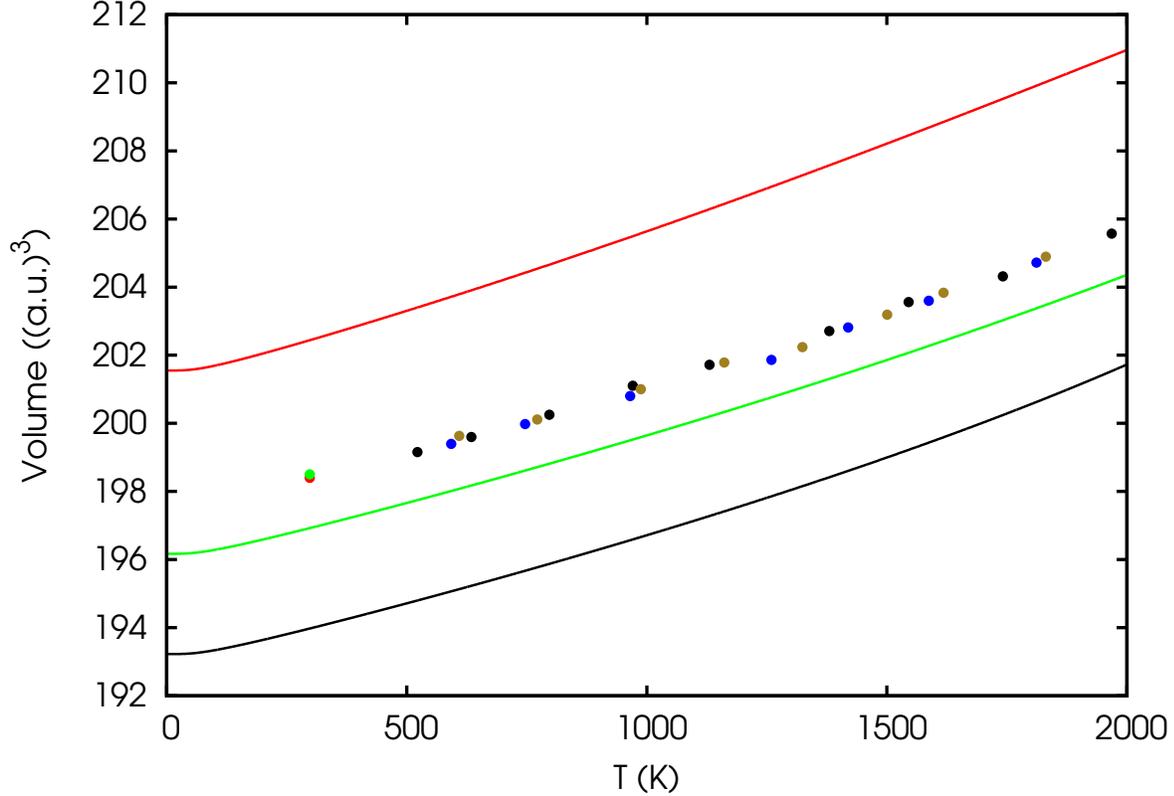}
\caption{Quasi-harmonic volume of h.c.p. Re calculated as a function of temperature with both LDA (black line), PBE (red line) and PBEsol (green line) exchange-correlation 
functionals. The calculations were done assuming a constant $c/a$ ratio and using the Murnaghan EOS.  
Experimental data at T=298~K are from Refs.\cite{87Voh,70Liu}. Thermal expansion data from~\cite{66Con} were used to derive
the volumes (points above 500~K) as a function of temperature.
}\label{fg:volume}
\end{figure}

As outlined in the previous section, anharmonic quantities were first calculated by fixing the $c/a$ ratio and using the Murnaghan EOS.
The volume as a function of temperature is reported in Fig.~\ref{fg:volume}.  It can be noted that the temperature dependence
of the volume is in remarkable agreement with the experimental data, but the LDA volume underestimates the experiments by $\approx2\%$.
The discrepancy is the typical limitation of the exchange-correlation functional, as proved by the PBE results which on the contrary overestimate
the experiments of approximately the same amount. As a further confirmation, the PBEsol functional, suited to reproduce the structural properties
of densely-packed solids, is in better agreement with experiments ($<1\%$).
For volumes higher than about 15.15 \AA$^3$ (and $c/a$ constant as in 
Tab.~\ref{tab:0K}) in the present LDA calculations 
the LO frequency at $\Gamma$ and H become
imaginary. Similarly, for the full anisotropic grid imaginary frequencies are obtained for high $a,c/a$ values. The occurrence
of these imaginary frequencies is peculiar of h.c.p. Re (and Tc, see next section), which becomes mechanically unstable~\cite{14Pal} 
when its unit cell is expanded, but not of other h.c.p. elements.
The volume values and $(a,c/a)$ grid have been carefully chosen in order to avoid the occurrence of imaginary frequencies which cannot
be used in the quasi-harmonic approximation. This limits the applicability of the quasi-harmonic approximation
to temperatures well below the melting point. We expect therefore that anharmonic effects beyond the quasi-harmonic stabilize h.c.p. Re
at high volumes/temperatures. PBE and PBEsol results show an even stronger 
softening effect and are thus not used to study other quantities in the following.

\begin{figure}[htp]
\includegraphics[height=\linewidth, angle=-90]{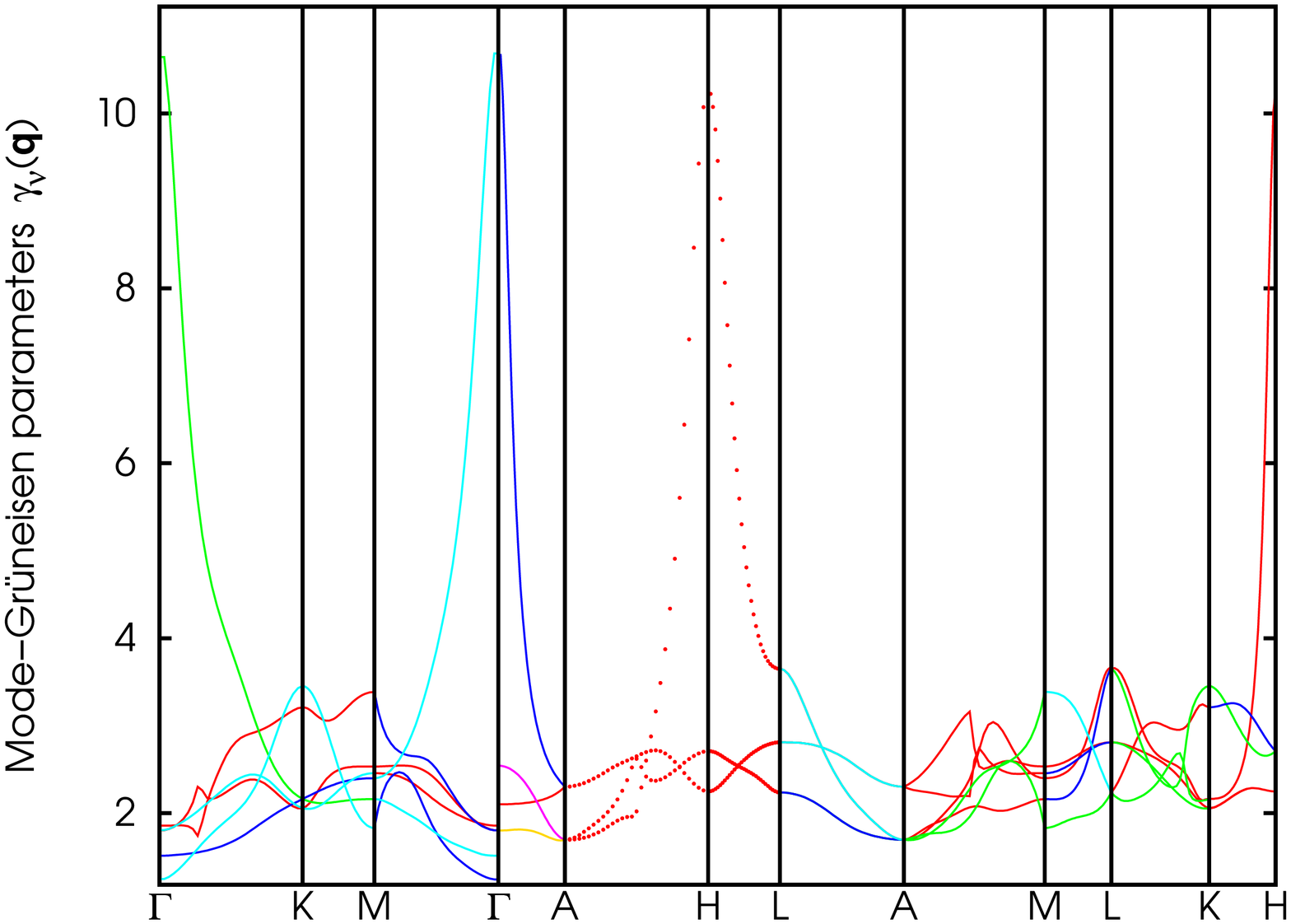}
\caption{Gr\"{u}neisein mode parameters of h.c.p. Re calculated as in Eq.~\ref{eq:Grun} with LDA exchange-correlation functional along high symmetry points in the BZ. 
}\label{fg:Grun}
\end{figure}

\begin{figure}[htp]
\includegraphics[width=\linewidth]{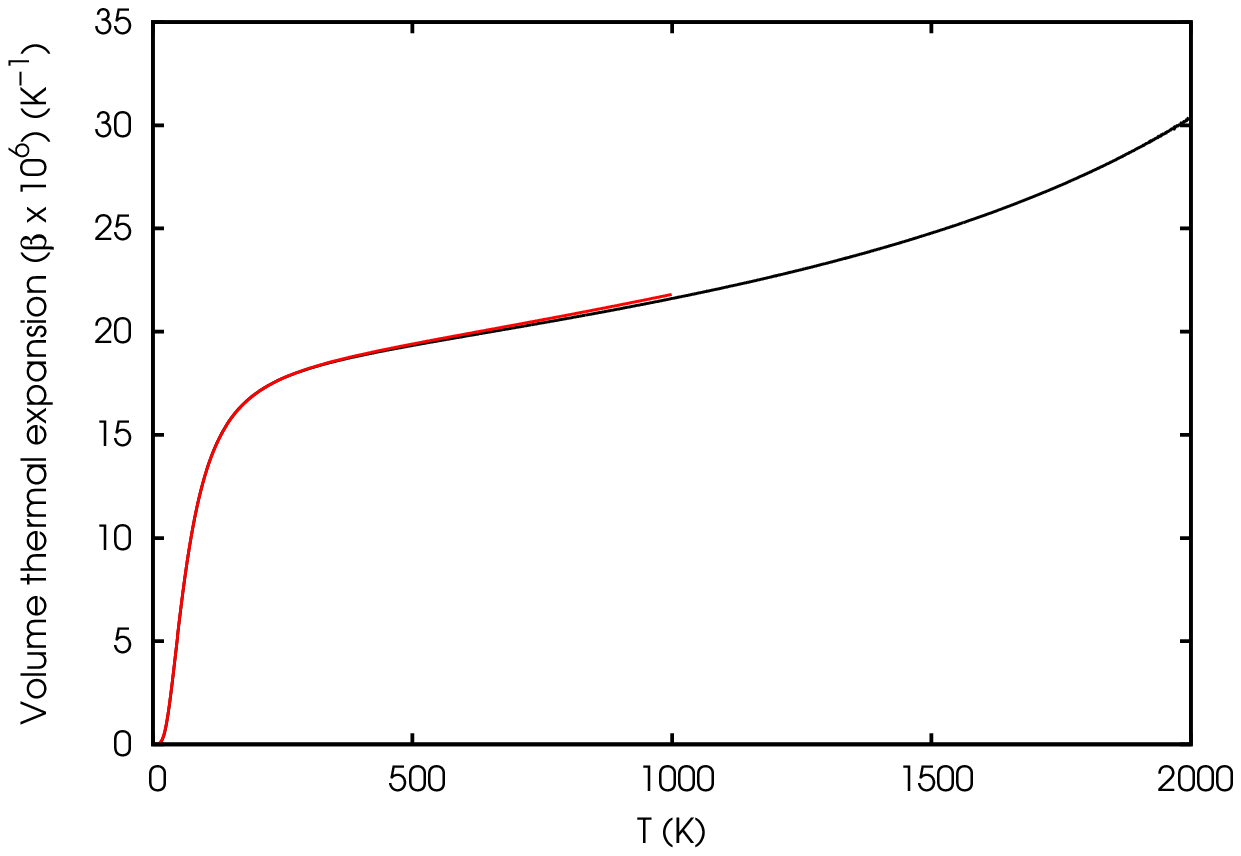}
\caption{Quasi-harmonic volume thermal expansion of h.c.p. Re calculated as a function of temperature with LDA exchange-correlation functional. The black line is obtained
assuming a constant $c/a$ ratio and using the Murnaghan EOS. The red line is obtained using the full anisotropic results for the linear thermal 
expansions ($\alpha_1$ and $\alpha_3$) and $\beta=2 \alpha_1 + \alpha_3$. The results in the anisotropic case are only up to 1000~K in order to keep a 
comparable fitting quality with the Murnaghan case.
}\label{fg:thermalexpansion}
\end{figure}

\begin{figure}[htp]
\includegraphics[width=\linewidth]{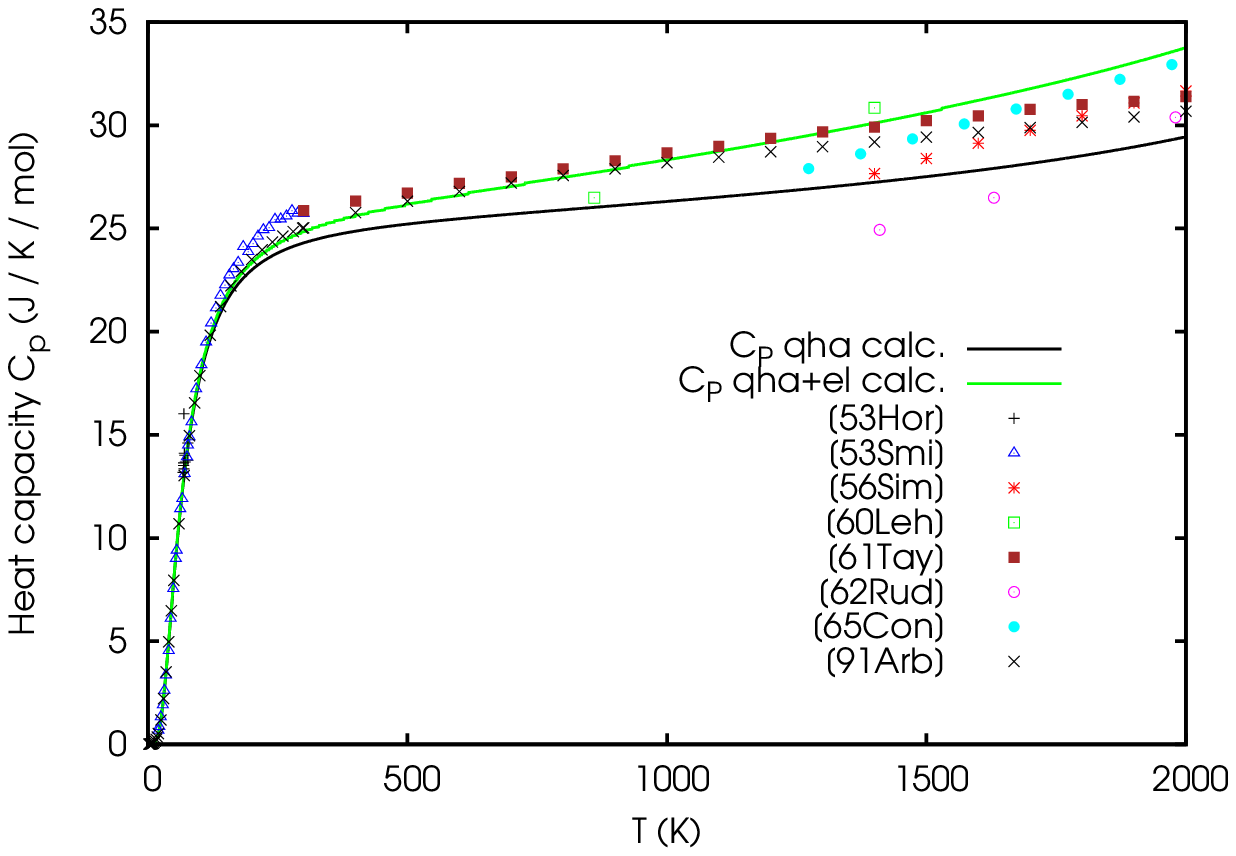}
\caption{Calculated isobaric heat capacity of h.c.p. Re as a function of temperature with LDA exchange-correlation functional, quasi-harmonic only ($C_P$ qha) and 
quasi-harmonic plus electronic ($C_P$ qha+el). The calculations were done assuming a constant $c/a$ ratio and using the Murnaghan EOS. 
Experimental data from different sources are shown as points: 53Hor~\cite{53Hor}, 53Smi~\cite{53Smi}, 56Sim~\cite{56Sim},
60Leh~\cite{60Leh}, 61Tay~\cite{61Tay}, 62Rud~\cite{62Rud}, 65Con~\cite{65Con}, 91Arb~\cite{91Arb}. Points from Ref.~\cite{91Arb} 
are not pure experimental data but assessed values obtained from a critical evaluation of available experimental data.
}\label{fg:Cpeldos}
\end{figure}

The Gr\"{u}neisein mode parameters at the 0~K equilibrium volume calculated as in Eq.~\ref{eq:Grun} are shown in Fig.~\ref{fg:Grun}. 
They are positive for all phonon modes and
branches and thus all frequencies in the Re phonon dispersion are expected to soften with increasing volumes. The magnitude of the Gr\"{u}neisein
parameters shows remarkable differences among the phonon modes and points in the BZ. Most of them lie in the range 1 to 3.5, but the LO mode 
parameters at both at $\Gamma$ and H are particularly sensitive to volume changes as already discussed.
The quasi-harmonic thermal expansion and isobaric heat capacity are shown in Fig.~\ref{fg:thermalexpansion} and Fig.~\ref{fg:Cpeldos}, respectively. 
In the latter, a significant number of experimental data is also shown. The points from Ref.~\cite{91Arb} are assessed values obtained from a critical 
evaluation of the available experimental data and they agree well with the calculated results when the electronic contribution is taken into account.

\begin{figure*}[htp]
\subfigure[]{
\includegraphics[width=.5\textwidth]{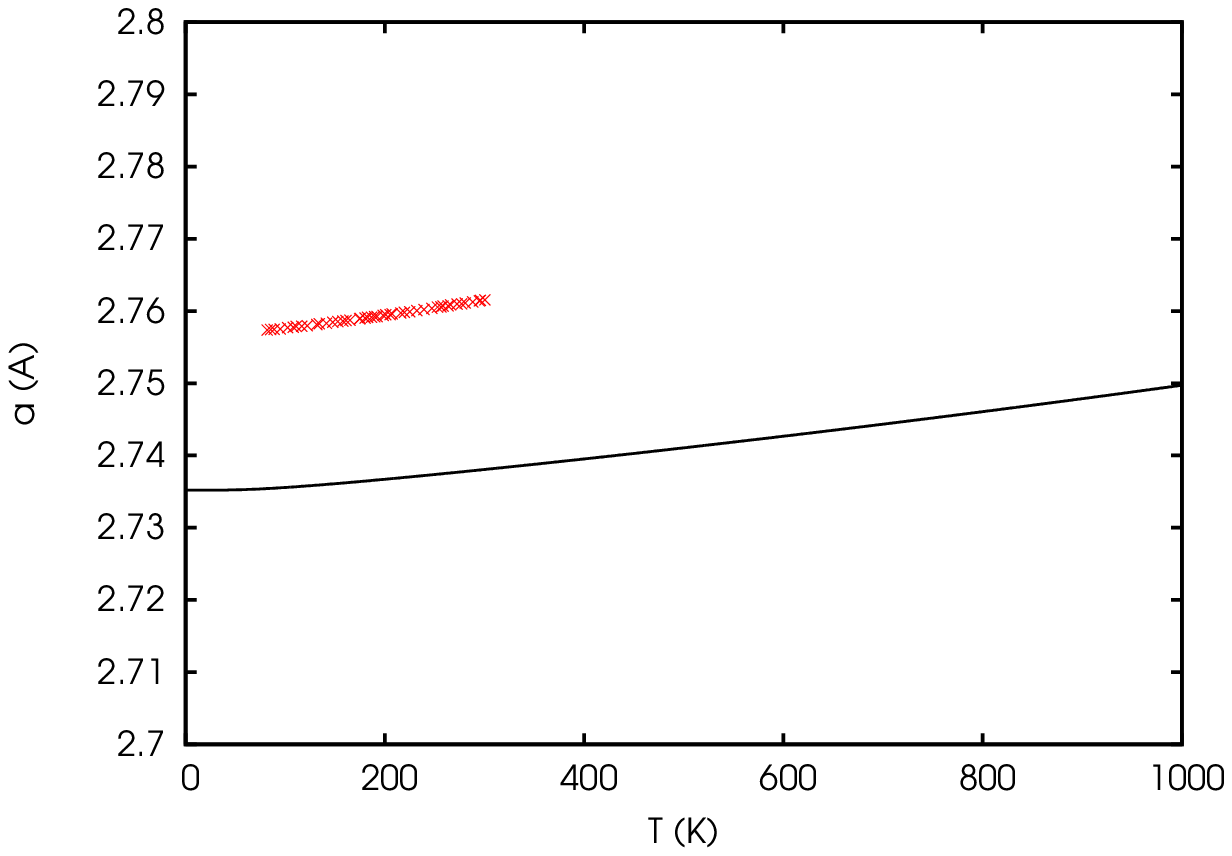}}
\subfigure[]{
\includegraphics[width=.5\textwidth]{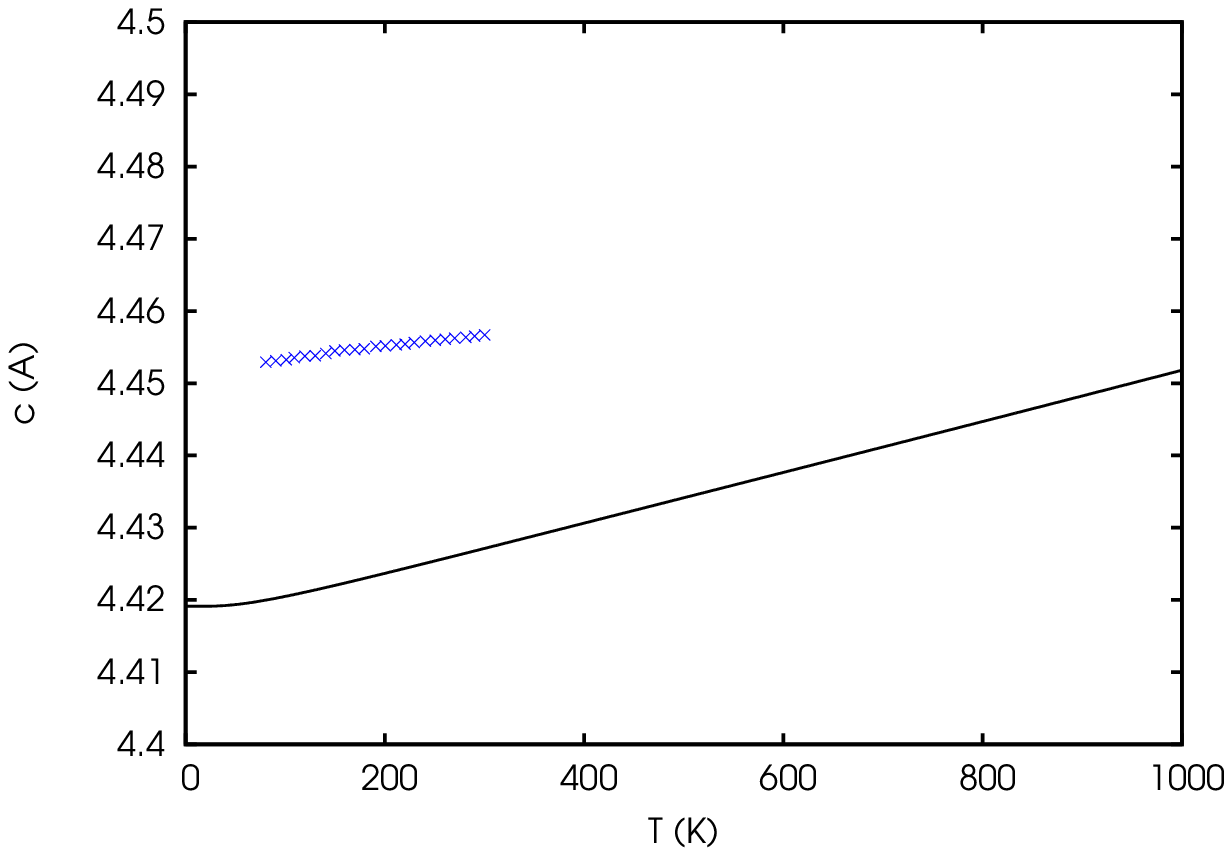}}
\subfigure[]{
\includegraphics[width=.5\textwidth]{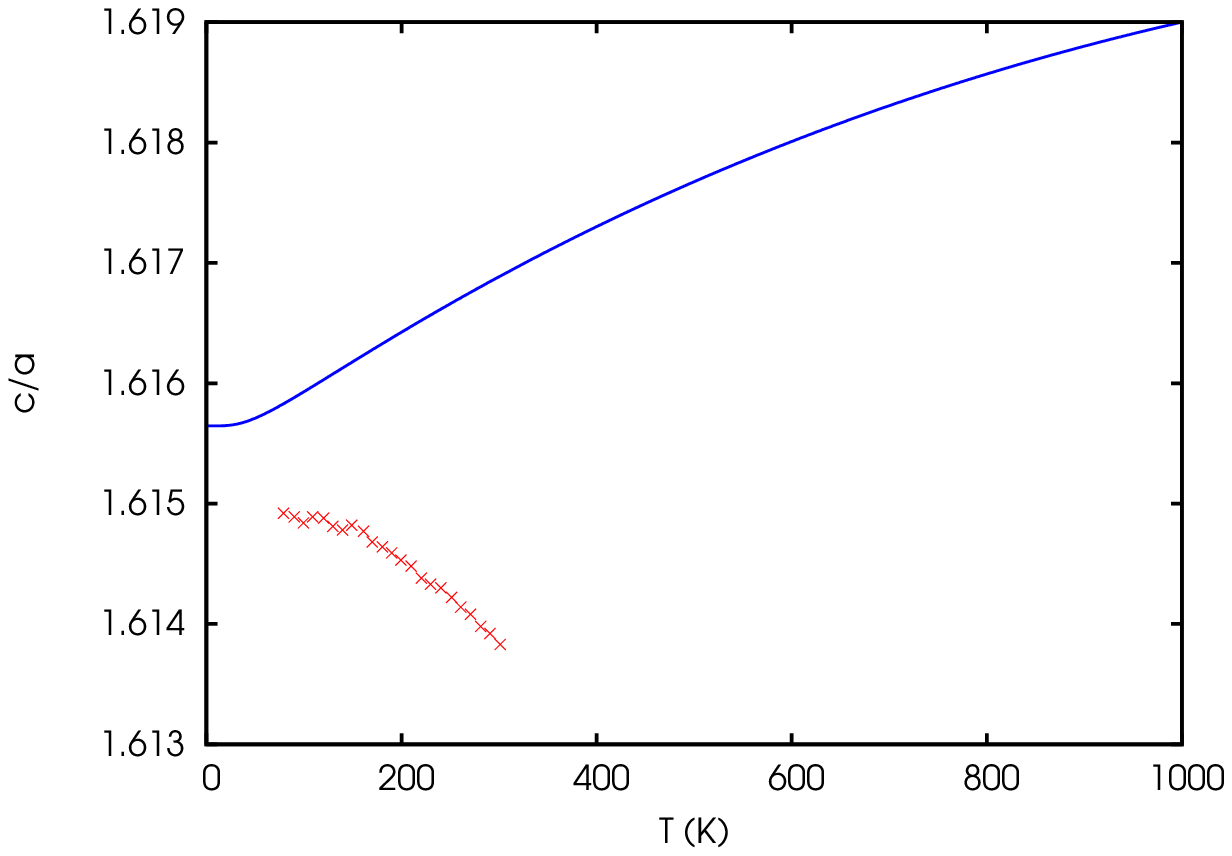}}
\caption{Quasi-harmonic variation of the lattice parameters $a$, $c$ and $c/a$ of h.c.p. Re compared with experimental points from Ref.~\cite{71Fin}.
These results were obtained using the full grid ($a$,$c/a$),
minimizing the Helmholtz energy to obtain $a(T)$ and $c/a(T)$
\label{fg:Relat}
}
\end{figure*}

\begin{figure}[htp]
\includegraphics[width=\linewidth]{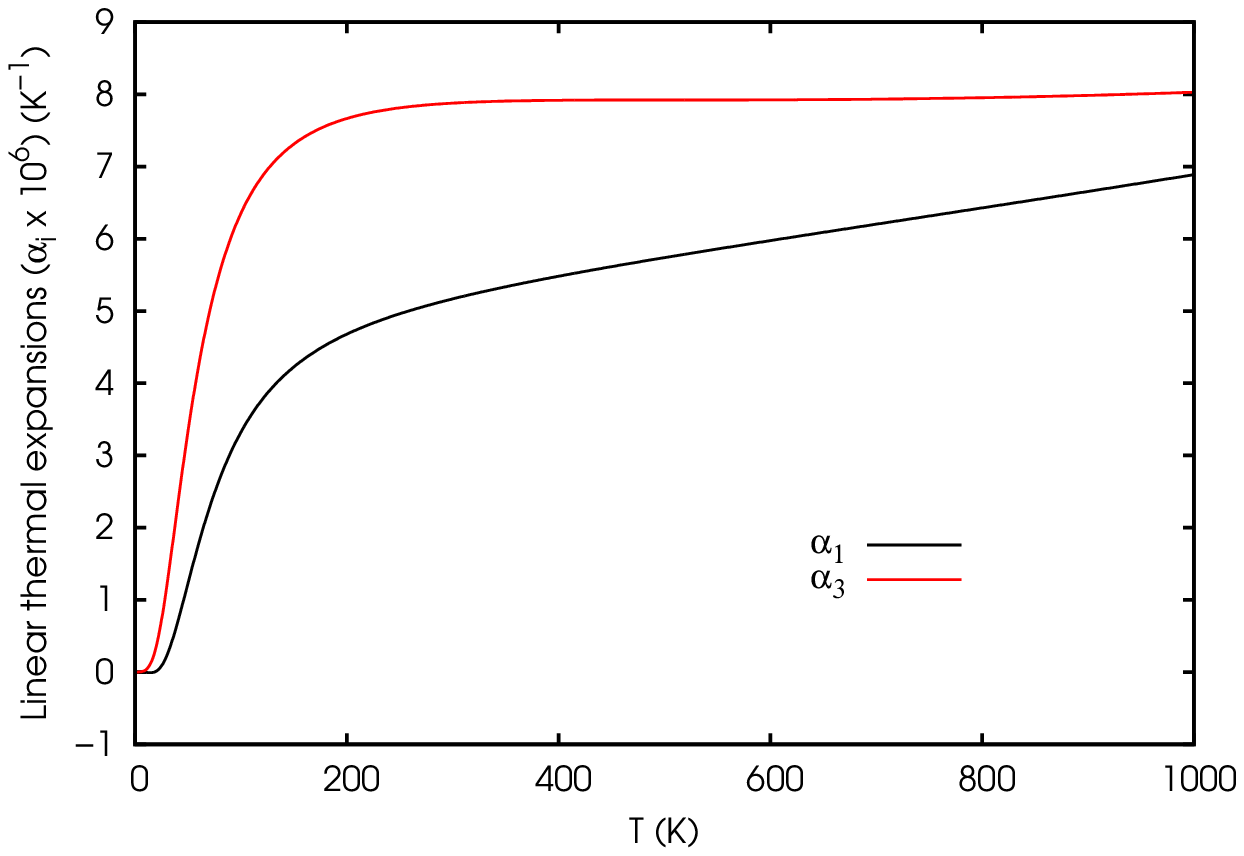}
\caption{Calculated LDA linear thermal tensor ($\alpha_1=\alpha_2$ and $\alpha_3$) for Re h.c.p. These results were obtained using the full grid ($a$,$c/a$),
minimizing the Helmholtz energy to obtain $a(T)$ and $c/a(T)$ and eqs.~\ref{alphaa},~\ref{alphac}.
}\label{fg:Re_alphas}
\end{figure}

We already remarked that the results hereto discussed rely on the assumption that the $c/a$ ratio does not vary significantly with temperature. In order
to verify this assumption and to explore the full anisotropic properties of h.c.p. Re, we carried out phonon calculations on a grid of $a$ and $c/a$ values
centered around the equilibrium values, as discussed in Section~\ref{sec:anisotropy}. In this way, the independent evolution with temperature of $a$, $c$ and $c/a$
can be determined from the minimization of the Helmholtz energy and the results are reported in Fig.~\ref{fg:Relat}. 
The anisotropic thermal expansion tensor is shown in Fig.~\ref{fg:Re_alphas} as calculated from eqs.~\ref{alphaa},~\ref{alphac}.
The temperature dependence of both $a$ and $c$ compare satisfactorily with experimental data, with the usual offset
due to the exchange-correlation error. It is worth noting that the temperature dependence of the $c/a$ is rather limited in the calculated results, it increases only by 0.3\% from 0~K
to 1000~K. This explains why the volumetric thermal expansion $\beta$ obtained from the full anisotropic calculation in Fig.~\ref{fg:thermalexpansion} combining 
$\alpha_1=\alpha_2$ and $\alpha_3$ ($\beta=2\alpha_1+\alpha_3$) is nearly identical to that obtained from the Murnaghan EOS calculation. The experimental measurements of $c/a$
reported by Frenkel et al.~\cite{71Fin}, however, show an opposite trend with temperature compared with our calculated results (Fig.~\ref{fg:Relat}). 
This behavior appears to be unusual and different from what was measured
by the same authors for other h.c.p. elements such as Os and Ru. 
The total variation in Frenkel's data for $c/a$ is ~1\% in the temperature range 80-300~K, higher than in the calculated results.

\subsection{Technetium}\label{secTc}

First, we compared the calculated results at 0~K for the structural parameters of h.c.p. Tc with available experimental data in Tab.~\ref{tab:Tc0K}. 
Our LDA values
underestimate the experimental data, a result which is both due to the exchange-correlation functional and the absence of temperature-dependent 
effects. In fact, the inclusion of the latter partially improves the agreement with experiments (see the following). We remark that the PBE results
from Weck et al. show a higher value for both $a$ and $c$, but a comparable value for the $c/a$ ratio with respect to ours.

\begin{table}[htp]
\caption{\label{tab:Tc0K} Comparison between structural parameters of h.c.p. Tc from different sources. 
The PAW LDA results in this work were obtained at 0~K using the Tc.pz-spn-kjpaw\_psl.1.0.0.UPF. Experimental data are at room temperature.}
\begin{center}
\begin{tabular}{@{}lcccc}
 Method & $a$ & $c$ & $c/a$ \\
        & $\AA$ & $\AA$ & \\
PAW, LDA (this work, 0~K) & 2.712 & 4.334 & 1.598  \\
PAW, LDA (this work, 298~K) & 2.718 & 4.349 & 1.600  \\
PAW, PBE~\cite{15Wec} & 2.756 & 4.398 & 1.596 \\
Exp.~\cite{48Moo}    & 2.735 & 4.388 & 1.604 \\
Exp.~\cite{61Lam}    & 2.743 & 4.400 & 1.604 \\
Exp.~\cite{64Mul}    & 2.7415 & 4.400 & 1.605 \\
Exp.~\cite{65Bak}    & 2.7414 & 4.3997 & 1.6049 \\
Exp.~\cite{62Trz}    & 2.740 & 4.398 & 1.605 \\
Exp.~\cite{68Koc}    & 2.743 & 4.400 & 1.604 \\
Exp.~\cite{72Mar}    & 2.7407 & 4.3980 & 1.6047 \\
Exp.~\cite{75Spi}    & 2.73 & 4.39 & 1.61 \\
Exp.~\cite{85Gio}    & 2.740 & 4.399 & 1.605 \\
Exp.~\cite{80Hai}    & 2.7375 & 4.3950 & 1.6055 \\
Exp.~\cite{99Rar}    & 2.7409 & 4.3987 & 1.6048 \\
\end{tabular}
\end{center}
\end{table}

\begin{figure}[htp]
\includegraphics[height=\linewidth, angle=-90]{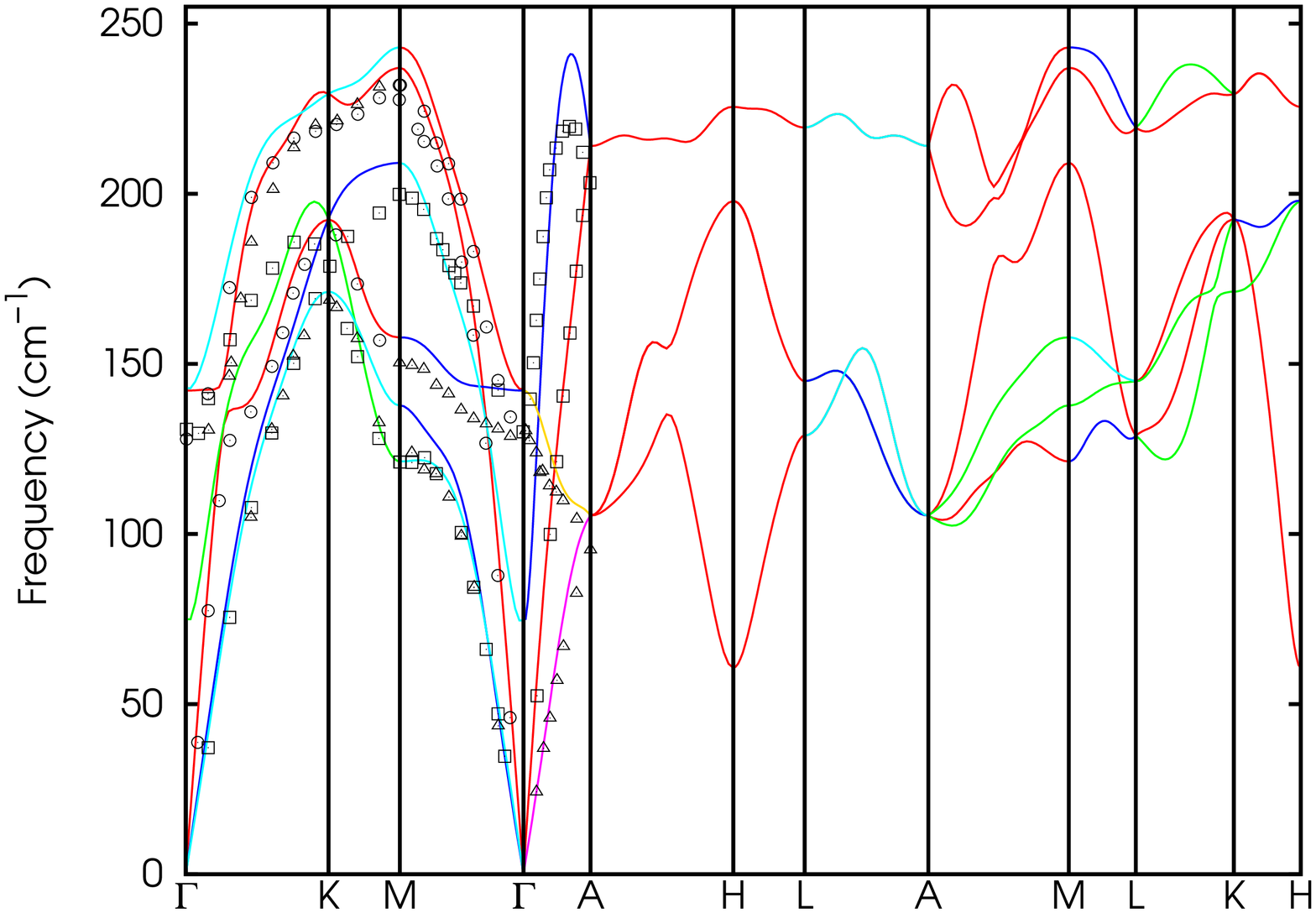}
\caption{Phonon dispersion curves for Tc along the same high symmetry points and directions in the BZ as for Re calculated using the 
LDA exchange-correlation functional. The calculation was carried out at the equilibrium lattice parameters at 298~K. Experimental data (points) are from Ref.~\cite{82Wak} at 298~K.
}\label{fg:Tcphdisp}
\end{figure}

As for Re, we computed phonon dispersions of h.c.p. Tc at 298~K ($a$=2.718 \AA, $c/a$=1.600) and the results are shown in Fig.~\ref{fg:Tcphdisp}. 
With respect to Re, more experimental
phonon frequencies are available and it can be noted that the calculated results are in satisfactory agreement
with the experimental points in most branches.
A slight discrepancy can be observed at the highest frequencies. We remark that, as for Re, soft modes are observed both at $\Gamma$ and at H. 
The mode Gr\"{u}neisein parameters for Tc (not shown) are similar to Re and in particular they present high values at $\Gamma$ and at H. Hence, 
the softening
of these modes when expanding the unit cell
is remarkable (even more pronounced than for Re) and imaginary frequencies appear in the phonon 
dispersions. 
In the rest of this work we limit our discussion for Tc to results up to approximately 1500~K (with the Murnaghan EOS) and 500~K (with the anisotropic grid),
where the fitting and minimization of the quasi-harmonic free energies can be done with good accuracy. We note that the quasi-harmonic approximation is 
commonly expected to give good results up to (roughly) $2/3$ of the melting temperature, which is in agreement with our findings for Re and Tc.

As already remarked, experimentally the LO mode frequency at $\Gamma$ increases with temperature for Tc, contrary to what is found in the quasi-harmonic
approximation. Using the Fermi smearing as discussed for Re, we have checked the electronic effect on the LO mode frequency for Tc, but contrary to Re,
the frequency (84 cm$^{-1}$) does not increase by varying the Fermi smearing  up to 600~K thus failing to reproduce the experimental trend in this case.

Similarly to Re, we carried out calculations of several thermophysical quantities for Tc, first assuming a fixed $c/a$ ratio (at the 0~K equilibrium value)
and using the Murnaghan EOS, then doing the quasi-harmonic calculations on a full $(a,c/a)$ grid.

\begin{figure}[htp]
\includegraphics[height=\linewidth, angle=-90]{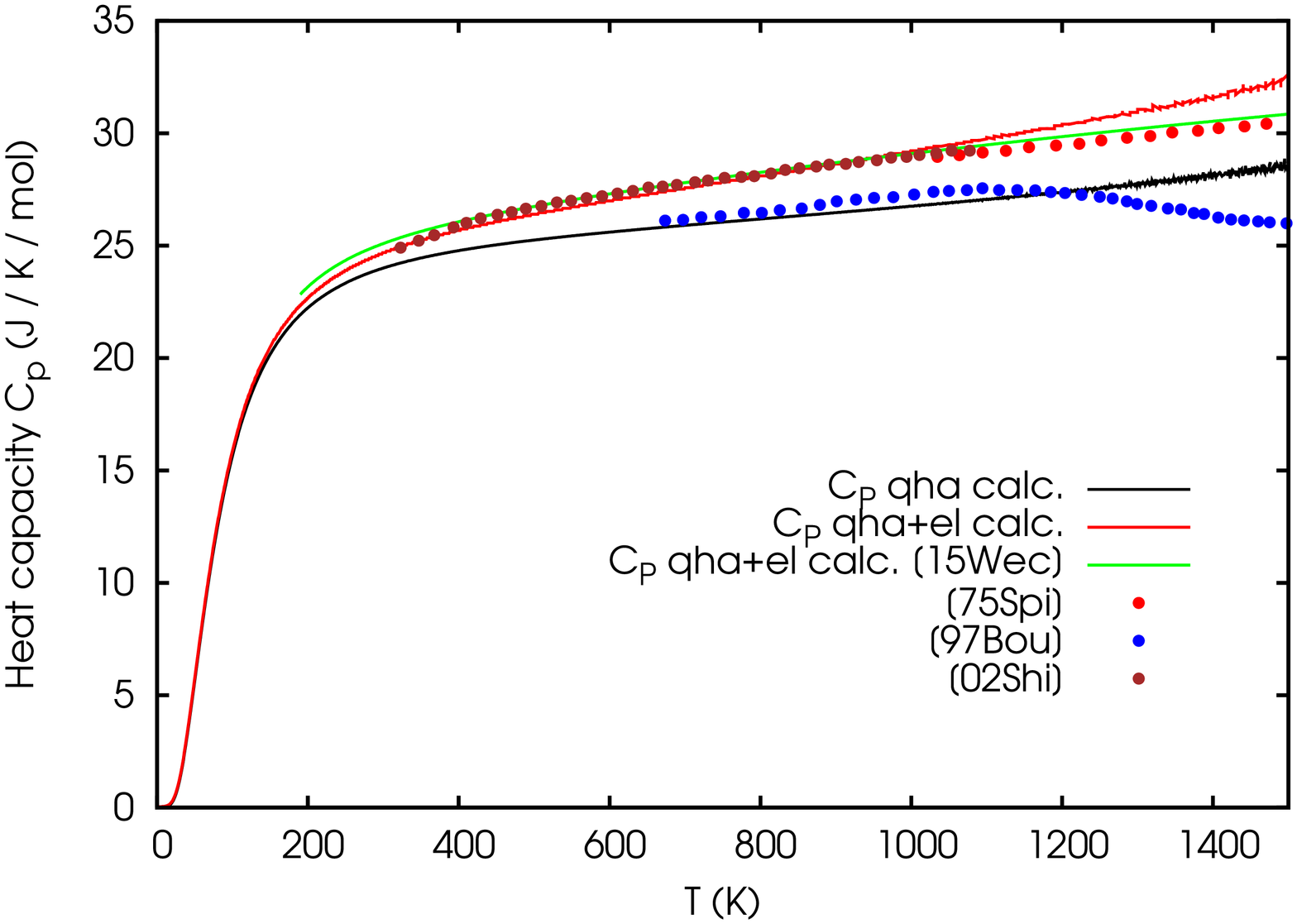}
\caption{Quasi-harmonic isobaric heat capacity for h.c.p. Tc calculated as a function of temperature (with and without electronic contribution)
with LDA exchange-correlation functional compared with experimental data (points). Experimental points are from Refs.~\cite{75Spi} (red points, 75Spi),
~\cite{97Bou} (blue points, 97Bou) and \cite{02Shi} (green points, 02Shi). The calculated values from Ref.~\cite{15Wec} are also shown for comparison 
(green line). This line was obtained using the Haas-Fisher polynomial coefficients reported in Tab.~2 of the original paper, in the range of validity of
the fitting (190-1620~K).
}\label{fg:TcCp}
\end{figure}

\begin{figure}[htp]
\includegraphics[height=\linewidth, angle=-90]{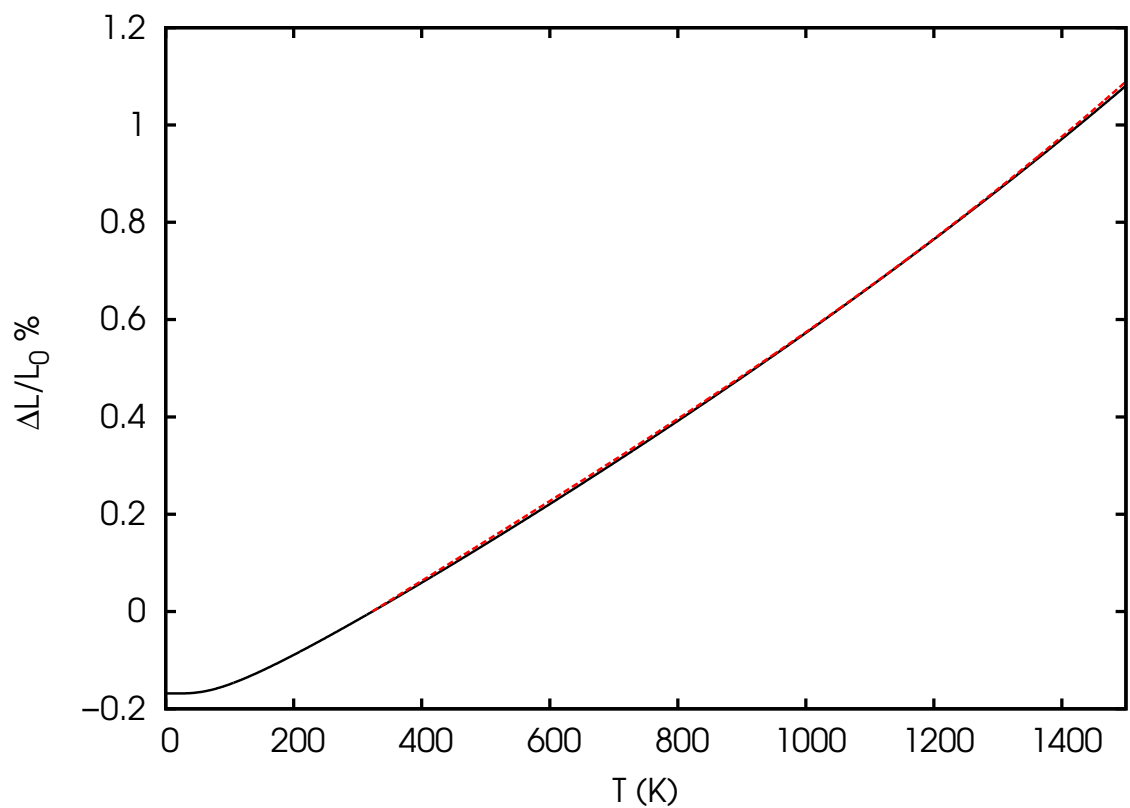}
\caption{Quasi-harmonic thermal expansion of Tc calculated as a function of temperature with LDA exchange-correlation functional using the Murnaghan EOS 
(black line). The red dashed line represents the experimental measurements as reported in Ref.~\cite{02Shi2},
i.e. fitted using a polynomial in the temperature range from 323 to 1300~K. 
For the sake of comparison, the calculated values are reported as $\Delta L/L_0$, with $L_0=V_0^{1/3}$ corresponding to T=323~K.
}\label{fg:Tcthermalexpansion2}
\end{figure}

Fig.~\ref{fg:TcCp} shows the results obtained for the heat capacity using the Murnaghan EOS, with and without the electronic contribution. As it can be
seen, the results compare well both with the experiments and with the previous results in Ref.~\cite{15Wec} when including the effect of electronic excitations. The lowest data set from~\cite{97Bou} is of
uncertain reliability, as already pointed out by Rard et al.~\cite{99Rar}. At high temperature, the presence of anharmonic effects beyond the quasi-harmonic
approximation can explain the increasing gap between the experiments and the calculations. An excellent agreement is obtained for the calculated 
thermal expansion compared with the experimental results from Ref.~\cite{02Shi2}, as shown in Fig.~\ref{fg:Tcthermalexpansion2}.

\begin{figure}[htp]
\includegraphics[height=\linewidth, angle=-90]{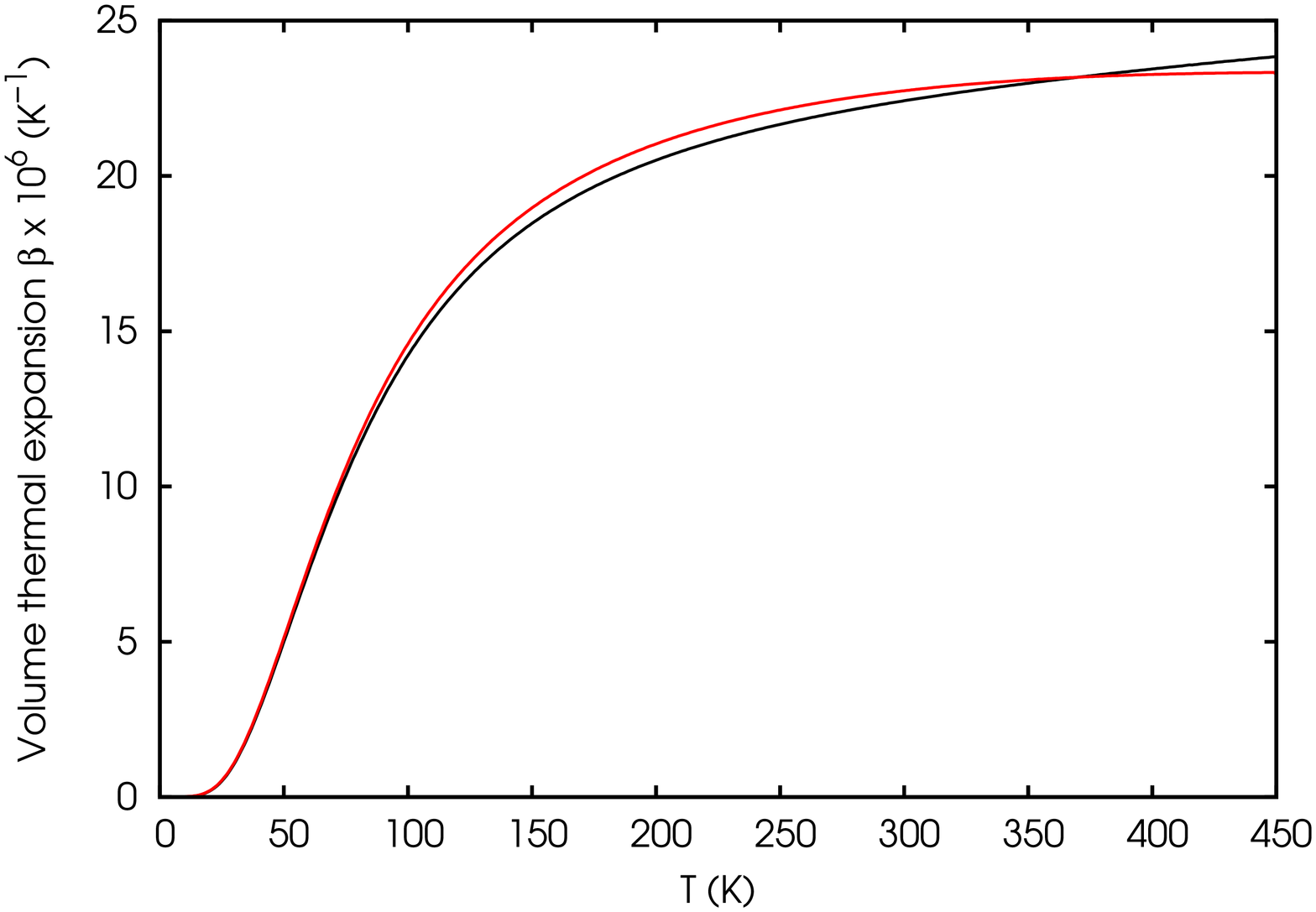}
\caption{Quasi-harmonic volume thermal expansion of Tc calculated as a function of temperature with LDA exchange-correlation functional. The black line is obtained
assuming a constant $c/a$ ratio and using the Murnaghan EOS. The red line is obtained using the full anisotropic results for the linear thermal 
expansions ($\alpha_1$ and $\alpha_3$) and $\beta=2 \alpha_1 + \alpha_3$. The results in the anisotropic case are limited up to 450~K in order to keep a 
comparable fitting quality with the Murnaghan case.
}\label{fg:Tcthermalexpansion}
\end{figure}

\begin{figure}[htp]
\includegraphics[height=\linewidth, angle=-90]{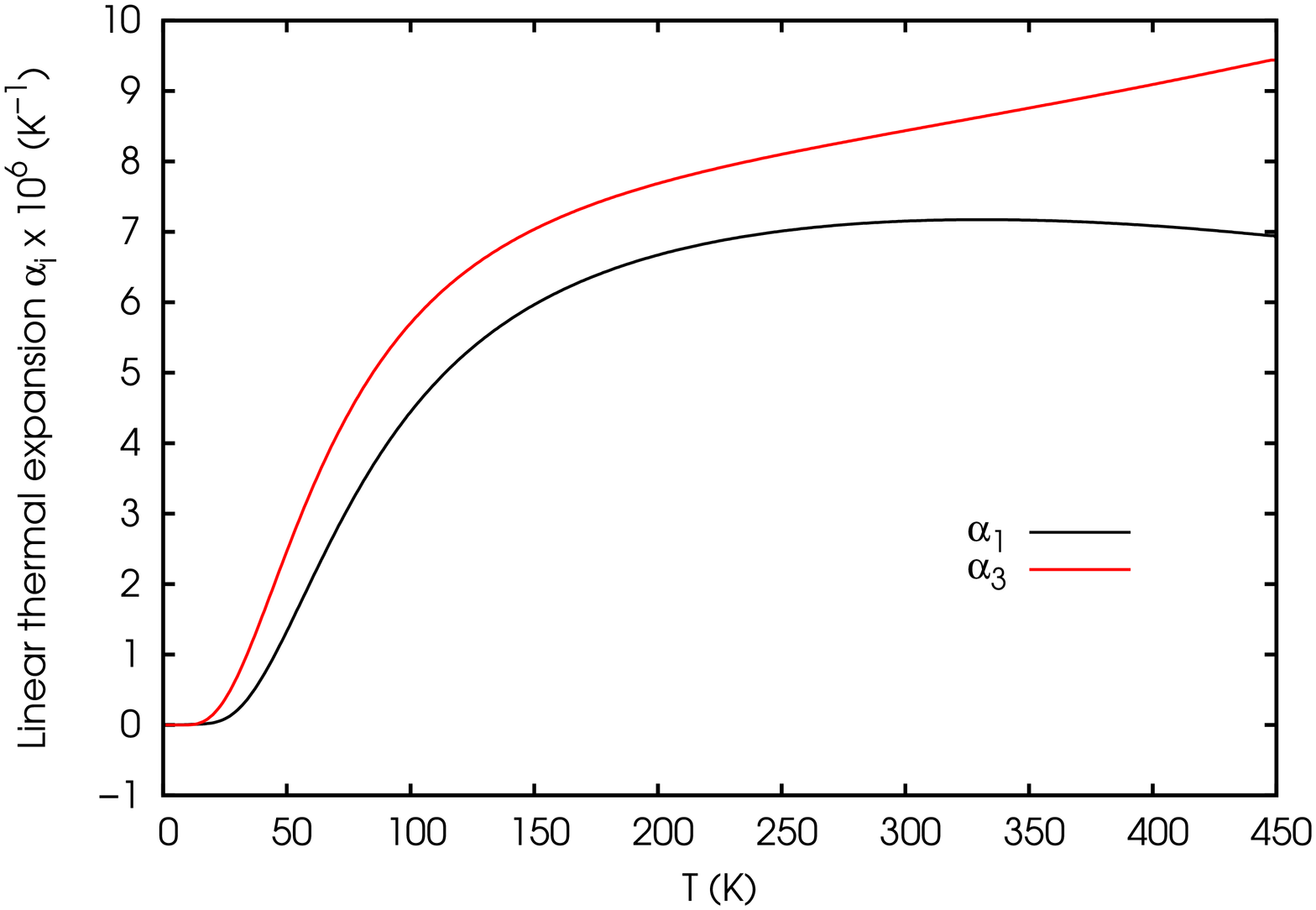}
\caption{Calculated LDA linear thermal tensor ($\alpha_1=\alpha_2$ and $\alpha_3$) for Tc h.c.p. These results were obtained using the full grid ($a$,$c/a$),
minimizing the Helmholtz energy to obtain $a(T)$ and $c/a(T)$ and eqs.~\ref{alphaa},~\ref{alphac}.
}\label{fg:Tcalphas}
\end{figure}

As for Re, the full anisotropic calculation gives isotropic quantities very similar to those obtained with the Murnaghan EOS and constant $c/a$ ratio.
The calculated $c/a$ ratio goes from 1.599 at 0~K 
(including ZPE) to 1.600 at 500~K. The anisotropic results of the volume thermal expansion are nearly identical to those obtained using a 
fixed $c/a$ ratio, as evidenced in Fig.~\ref{fg:Tcthermalexpansion}. The anisotropic linear thermal expansions are shown in Fig.~\ref{fg:Tcalphas} and 
are closer to each other than for Re, in agreement with the nearly constant $c/a$ ratio found.
Nonetheless, we have to remark that in the present case the highest 
temperature we could reach in the anisotropic results is rather limited.

\section{Conclusion}
In this work, the lattice dynamics and thermophysical properties of h.c.p. Re and Tc were investigated from first-principles 
using density functional theory and the quasi-harmonic approximation. The calculations were carried out both assuming 
a temperature independent $c/a$ ratio and fully analyzing the anisotropic properties of both elements. In the latter case, the temperature
dependence of the structural parameters and the thermal expansion tensor were determined.

Both Re and Tc present some peculiar features in their phonon dispersions with respect to other h.c.p. elements,
i.e. the presence of soft modes at the $\Gamma$ and H
points in the BZ. These experimental findings have been confirmed by the present calculations. Besides,
the phonon frequencies at these two points are strongly volume dependent and imaginary frequencies occur when expanding the unit cell. This
happens for volumes below the melting point suggesting the presence of stabilizing anharmonic effects beyond the quasi-harmonic approximation.
The calculated phonon frequencies are in satisfactory agreement with the experimental data available for both Re and Tc. The increase with temperature
of the experimental frequency of the LO mode at $\Gamma$ in Tc is however in contradiction with the quasi-harmonic results and it is not fully explained
by electronic effects which have the correct trend but strongly underestimates the magnitude of the experimental effect.

The thermal properties of both h.c.p. Re and Tc obtained with the Murnaghan EOS and a constant $c/a$ ratio agree
well with the experimental data when available. For the heat capacity, the agreement with experiments requires the inclusion of 
the electronic excitations contribution, which in not negligible. For Tc, our results also agree well with previous findings by Weck et al. 
The present results for Tc also confirm doubts on the reliability of the heat capacity experimental data in Boucharat's thesis. 

The anisotropic results provide the temperature-dependent lattice parameters and thermal expansion tensor and they agree satisfactorily with the available
experimental data. The anisotropic results show
that the calculated $c/a$ ratio is nearly constant in the temperature range investigated. Thus it is not 
surprising that the volume thermal expansion $\beta$ obtained from the full anisotropic tensor $\alpha$ is very close to that obtained
from the Murnaghan EOS with constant $c/a$ ratio. It remains to be seen if this result is true for other h.c.p. systems. Experimentally, however, a slight 
decreasing trend with temperature, not reproduced by our calculations, was found for the $c/a$ of Re, while no data are available for Tc.
The main discrepancy between calculated and experimental results can be ascribed to the exchange-correlation functional 
in the temperature range where data are available for the two present elements and well below the melting point. 
However, the quasi-harmonic approximation cannot
be applied up to the melting point for the occurrence of imaginary frequencies. The inclusion of full anharmonic effects
would be necessary to this aim.

\section*{Acknowledgement}
This work was performed within the MaX Center of Excellence, with support from the European Union
Horizon 2020 EINFRA program under grant agreement No 676598. Computational facilities have been provided by SISSA through its Linux
Cluster and 	ITCS.

\section*{References}

\bibliographystyle{unsrt}

\end{document}